# Spreadsheet Modeling Experiments Using GPTs on Small Problem Statements and the Wall Task

Thomas A. Grossman, Yuan Chen, and Sopiko Datuashvili
University of San Francisco
2130 Fulton St.
San Francisco, CA 94117 USA
tagrossman@usfca.edu, ychen462@dons.usfca.edu, sdatuashvili@dons.usfca.edu

## Abstract

This paper investigates how GPT-based tools can assist in building reusable analytical spreadsheet models. After a screening, we evaluate five GPT extensions, and select Excel AI by pulsrai.com for detailed testing. Through structured experiments on simple problem statements, we assess Excel AI's performance against the ERFR criteria (each input in a cell; cell formulas; no hardwired numbers; labels; accurate). Results show that while Excel AI can produce well-structured models, it is inconsistent and often non-reproducible. We identify two central challenges—"the problem of confidence" and "the problem of workflow"—which highlight the need for skilled users to verify and adapt GPT-generated spreadsheets. Though GPTs show promise for generating draft models that may reduce development time or lower skill requirements, current tools remain unreliable for professional use. We conclude with recommendations for future research into prompt engineering, reproducibility, and larger-scale modeling tasks.

## 1. INTRODUCTION

It is widely understood that Artificial Intelligence general pre-trained transformers (GPTs) are poised to have a large impact on many aspects of business activity. Analytics is perceived to be a particularly attractive candidate for GPTs, including spreadsheet analytics. A quick search of the internet reveals countless resources purporting to show how GPTs can help with analyzing datasets; visualizing data; devising cell formulas; and other useful activities. There are clearly some successes, although reliability remains an open question.

We seek to understand the reality of how GPTs can assist in spreadsheet analytics today, particularly in the creation of spreadsheet models. Our interest is the creation of reusable spreadsheet models that take input values and calculate output values. Our ultimate desire is "industrial quality analytical spreadsheet models" (Grossman, Mehrotra and Sander 2011; Grossman and Mehrotra 2023) which meet six criteria: suitable for efficient analysis; readable; transferable; accurate; reusable; modifiable. Because these criteria are subtle and complex, we will instead investigate a simpler goal, of reusability.

### 1.1 Essential Requirements for Reusability (ERFR)

For this research, we want to create spreadsheets that are reusable. We propose that there are five essential requirements for reusability (ERFR): 1) each input value resides in a single cell; 2) all calculations are done using cell formulas; 3) all calculations are done without hardwired numbers; 4) input and formula cells have appropriate labels; 5) the calculations are accurate. The first four requirements are necessary to ensure that the spreadsheet can be reused; they are largely mechanistic and can easily be confirmed by minimally-trained personnel. The fifth requirement of accuracy is necessary for the spreadsheet to be of practical value. However, as discussed below, confirming accuracy raises complex issues.

## 1.2 Contribution

We articulate the goal of automatic creation of industrial-quality analytical spreadsheet models from a prose problem statement. As an intermediate goal, we introduce the concept of essential requirements for reusability (ERFR): (each input in a cell; cell formulas; no hardwired numbers; labels; accurate) as the standard for evaluation of GPT spreadsheet output. To explore the capability of GPTs that claim to produce spreadsheet models, we performed a comparison of five GPTs on 16 simple prompts. We found that one GPT (Excel AI by pulsrai.com, "EAI") performed better than the others, and chose it for further study. We establish that EAI can often produce good spreadsheet models from simple problem statements.

We propose to structure prompts as a problem statement (that is platform and software-independent) plus a set of instructions (that is customized for each platform or software). We established a parsimonious set of three instructions for EAI. We examined three questions using eight short problem statements and found that EAI can (with caveats) handle made-up nouns, can appropriately model a month as 30 days, and can respond to problem statements that indicate data values as well as problem statements that are parameterized by indicating variables without values. We examined a paragraph-long problem statement (the Wall Task) and found performance to sometimes be excellent and sometimes poor.

We encounter issues of reproducibility (the same prompt produces different results on separate runs) and reliability (sometimes the output is good, sometimes poor). We raise the problem of confidence (how does one handle a spreadsheet provided by an agent, whether human or machine) and the problem of workflow (how might an imperfect spreadsheet model be helpful) and discuss their possible resolution.

Overall, we find that the goal of having a GPT produce industrial-quality analytical spreadsheet models is at best aspirational.

## 1.3 Overview

The paper is in six sections. Section 1 contains the introduction, ERFR criteria, and contribution statement. Section 2 explains the selection of GPT extensions for testing and evaluating. In Section 3, we explore the selected tool's performance and evaluate its ability to build ERFR spreadsheets. We extend our experiments to a larger problem statement, the Wall Task, and discuss the tool's performance in Section 4. To understand what might be the value of a GPT for building a spreadsheet model, we introduce the "problem of confidence" and "problem of workflow," in Section 5. Finally, Section 6 concludes and suggests future research.

# 2. SELECTION OF A SPREADSHEET MODELING GPT

There are any number of GPTs that claim to assist with spreadsheet analytics, including data analysis, data visualization, writing cell formulas, and creating spreadsheet models. We did not attempt to catalog them. We did an initial screen to narrow the field to five readily-available GPTs that claim to produce spreadsheet models. We used 16 small problem statements to evaluate performance, and identified one GPT that performed best; this GPT was then used for detailed testing.

## 2.1 Initial Screening

We selected five GPTs (Appendix 1) for initial screening based on the following criteria: 1) High Usage Volume: Each extension has recorded a minimum of 100,000 user conversations, indicating substantial user engagement and practical adoption. 2) Widespread Adoption and Popularity: All selected GPT extensions have received over 1,000 user reviews, suggesting they are widely used and recognized within the community. Additionally, each maintains a user rating of at least 3.3, reflecting generally favorable feedback. 3) Relevant Capabilities: All five extensions include Code Interpreter functionality, which is essential for supporting computational tasks and writing spreadsheet formulas. 4) Spreadsheet Modeling Capability: Each extension demonstrates the ability to generate valid, usable spreadsheet files (XLSX).

## 2.2 Evaluation

We evaluated GPT performance using a simple protocol.

**Evaluation Criteria**

We used the following five dimensions to evaluate the GPTs.

1. Conversation Quality - Refers to how detailed, informative, and easy to understand the explanations are during the interaction (e.g., in GPT chat). High conversation quality includes step-by-step guidance, clear communication of concepts, and explanations of formulas and calculations.
2. User Interaction - Measures how well the tool engages with the user, whether it asks clarifying questions, adapts to feedback, and guides the user through the process. Strong interaction also includes offering suggestions on how to improve the model or what features could be added.
3. Error Handling - Assesses how accurately the tool builds the model and how it handles mistakes. Good error handling results in fewer errors, and when errors do occur, they are easy to identify and correct thanks to a clear structure and step-by-step guidance in conversation chat.
4. File Reliability - Evaluates the quality and functionality of the generated spreadsheet files. A reliable file downloads correctly, opens without corruption in Excel, and includes working formulas and proper formatting with minimal issues.
5. Spreadsheet Design - Describes how well the spreadsheet is structured. A modular structure clearly separates inputs and outputs, labels provide meaning, and consistent formats make the model easy to read and understand.

**Evaluation Protocol**

We experimented using 16 short prompts to assess how different GPT tools responded to variations in prompt structure and detail. This allowed us to evaluate each model's prompt handling ability, as well as the impact of input formulation on output quality. We followed a consistent testing protocol for each GPT extension.

We entered a prompt and allowed the GPT to generate its response. If a downloadable file was not automatically provided, we added a follow-up prompt such as "provide a downloadable Excel file" to encourage file generation. If the GPT responded with follow-up questions or clarification requests, we answered those questions as a user would. We downloaded the Excel file and saved the chat history, preserving both the GPT's response and the user-GPT interaction for post-evaluation.

## 2.3 Selection of Excel AI by pulsrai.com

The results revealed significant variation in responsiveness, interaction depth, and structural consistency across the five GPTs. We selected "Excel AI" (EAI) by pulsrai.com because it had the best performance on the criteria above.

**Performance of EAI on Test Suite**

EAI consistently generated well-organized spreadsheet models that included clearly labeled input and output sections, accurate formulas, and reasonable formatting. Essential components such as model titles, variable input fields, step-by-step calculations, and final results were reliably present. These features were less prevalent in the other GPTs.

In contrast to some tools that produced broken or non-functional outputs, Excel AI consistently delivered clean and downloadable Excel files. These files contained valid and editable formulas along with structurally sound spreadsheets that were ready for immediate use. EAI showed the lowest incidence of technical errors among all tools evaluated. It consistently avoided common issues such as broken cell references, circular formulas, and inconsistent measurement units.

**Accessing EAI**

EAI is a "custom GPT" that can roughly be thought of as an add-in to ChatGPT. One can access EAI in ChatGPT by clicking "GPTs" in the left menu column, and searching on "Excel".

When using EAI, it is easy to inadvertently use regular ChatGPT without invoking Excel AI, in part because when one clicks "New Chat" in the left menu column, ChatGPT uses the default GPT rather than EAI; it is necessary to click "New Chat" under "Excel AI" at the top, just right of the left menu column (see Figure 1).

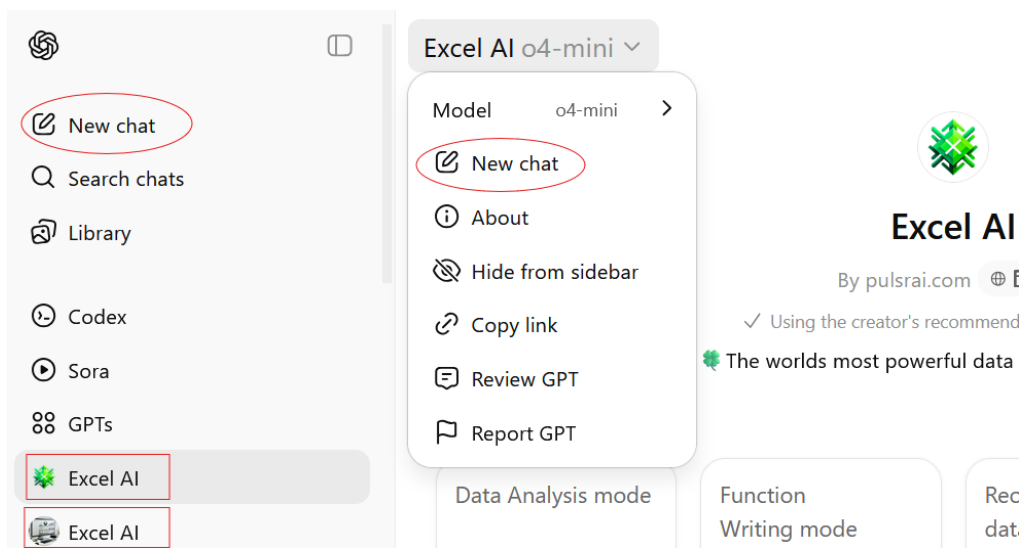

Figure 1. Two "New chat" options are in red ovals. Two GPTs named "Excel AI" are in red rectangles.

To increase confusion, there exist at least two custom GPTs named Excel AI (see Figure 1). To add further vexation, we note that searching the internet for "Excel AI" returns many things, none of which are EAI.

EAI recommends using ChatGPT o4-mini. However, we find that for our spreadsheet modeling goals, ChatGPT 4o is superior. The reason is that o4-mini likes to produce output in the form of images; these are painfully slow to build and require us to re-prompt to get an Excel file. Therefore, we use EAI with ChatGPT 4o.

More information on EAI can be found in Appendix 2.

# 3. SPREADSHEET MODELING EXPERIMENTS

We performed a series of experiments using EAI to evaluate its ability to build ERFR spreadsheets. We first discuss our prompting approach, and then describe a series of GPT runs.

## 3.1 Distinguish Between Problem Statement and Instructions

A prompt is the set of information we provide to the GPT. For our goal of getting the GPT to produce ERFR Excel files, it is helpful to distinguish between *problem statement* (what is to be calculated) and *instructions* (what is the desired output). Thus,

> Problem Statement + Instructions = Prompt

The *problem statement* is the pure logic of the question we are asking. A problem statement can be thought of as a "word problem" in a math textbook, or as a simple software specification.

There is an infinitude of reasonable outputs in response to a problem statement. The *instructions* are guidance as to the output we want in terms of *nature* (what we seek) and *form* (how it is shared). As to its nature, we might want the answer (a single number); a series of numbers that lead to a final number; a set of equations; or something else (e.g., a prose description of how to solve the problem). As to its form, we might want the answer presented as text of values or cell formulas; as an image displaying values or cell formulas; as an Excel file containing values or cell formulas; or something else (e.g., a poem, or perhaps an interpretive dance).

In the absence of clear instructions, a GPT (or for that matter, a human being) will necessarily make assumptions as to the nature and form of the output. Hence, it is incumbent upon the user to provide clear instructions. Indeed, it is unreasonable to expect the desired results if clear instructions are not provided.

**Three Instruction Statements**

We found during the screening process that the GPTs would sometimes present text, or create an image, rather than a spreadsheet file. To counteract this, we would interactively prompt with "Create an Excel model.". The Excel model might be displayed as formulas that we were encouraged to paste into Excel. To handle this, we interactively prompt "Provide a downloadable Excel file." The downloadable Excel file might have only numbers and no cell formulas. To counteract this, we interactively prompt "Use cell formulas".

After some experimentation, and in the spirit of parsimony, we found that the following three instruction statements seemed to result in good performance:

1. Create an Excel model.
2. Use cell formulas.
3. Provide a downloadable Excel file.

We make no claim that these are somehow the best, or even good, instructions. However, they did result in the GPT producing downloadable Excel files with cell formulas. (Caveat: additional work that is not presented in this paper, finds that these instructions are not as effective for larger models.)

## 3.2 Experiments

We established during the selection process that EAI can generally produce good spreadsheet models from simple prompts. Now we want to examine its abilities more systematically. We devised eight problem statements (Appendix 3) and used them to explore 3 questions:

1. How does the GPT perform when given Data v. Parameters?

   The distinction is between a problem where the variable values are presented, and when variable names without values (parameters) are presented. We compare Problem Statements 1 and 2 for a simple situation, and Problem Statements 3 and 4 for a slightly less simple situation.

2. How does the GPT handle "30 days" v. "month"?

   The distinction is between an explicit statement of 30 days, and an implicit indication of 30 days by stating a month. (Although it is a common analytic approximation to treat months as having 30 days, there is ambiguity because a month can have 28, 29, 30, or 31 days.) We compare Problem Statements 3 and 5 (for a prompt using Data), and Problem Statements 4 and 6 (for a prompt using Parameters).

3. How does the GPT handle a known noun v. a made-up noun?

   This distinction here is between a known noun and a made-up word that functions as a noun. We compare Problem Statements 3 and 7 (for a prompt using Data), and Problem Statements 4 and 8 (for a prompt using Parameters).

A prompt was created by appending the three instruction statements (section 3.1) to a problem statement (Appendix 3). For example, Problem Statement 1 is:

   Calculate the square footage of a rectangular room that is 10 feet long and 12 feet wide.

The corresponding prompt is:

   Calculate the square footage of a rectangular room that is 10 feet long and 12 feet wide. Create an Excel model. Use cell formulas. Provide a downloadable Excel file.

## 3.3 Results

We summarize the findings for each question. (Recall that each prompt consists of the problem statement plus three instruction statements.)

**How does the GPT perform when given Data v. Parameters?**

This is a comparison between Problem Statements 1 v. 2; and 3 v. 4. All four problem statements produced correct ERFRs.

For Problem Statements 1 and 2 (square footage), there were trivial differences in labels and the way units of measurement were indicated. We note that in Problem Statement 2, the two input cells were blank, and when we entered numbers into those cells the results were correct.

For Problem Statements 3 and 4 (a month of apples), the results were similar. There were trivial differences in labels. We note that in Problem Statement 4, the two columns of input cells were blank, and when we entered numbers into those cells the results were correct.

**How does the GPT handle "30 consecutive days" v. "month"?**

This is a comparison between Problem Statements 3 v. 5; and 4 v. 6. There was a failure here.

Problem statements 3 and 4 were shown to be correct ERFRs in the comparison above. Problem statement 5 produced a correct ERFR. Problem statement 6 failed. This failure is informative and is discussed in the next section.

**How does the GPT handle a known noun v. a made-up noun?**

This is a comparison between Problem Statements 3 v. 7; and 4 v. 8. All four problem statements produced correct ERFRs.

Problem Statements 3 and 7 (different nouns, problem statement uses data) are identical except for trivial differences in labels. Problem Statements 4 and 8 (different nouns, problem statement uses parameters) are identical except for trivial differences in labels. The made-up noun did not cause any issues.

**Summary**

Performance on the eight simple prompts is overall very good, but not perfect. EAI seems indifferent to being presented with parameters (variable names) as opposed to data examples. When data is not available, it either provides blank cells ready for data entry, or makes up representative data. EAI is aware that a month has 30 or 31 days. EAI has no problem handling a made-up noun.

However, one of the eight prompts resulted in a spreadsheet model that was accurate, but not useful because it had a column of numbers that needed to be a column of cell formulas. Although accurate in the sense that it correctly performed a calculation, this fails to meet the standard of ERFR because it is not reusable by changing inputs and having the outputs automatically update.

(Note: Out of curiosity, we ran all eight prompts in native ChatGPT-4o (without using EAI). All the spreadsheet models were correct, although they contained much less detail.)

## 3.4 The Failed Spreadsheet Model

Problem Statements 4 and 6 (Appendix 3) don't provide data and instead use parameter descriptions. They differ in that Problem Statement 4 indicates "30 consecutive days" and Problem Statement 6 indicates "for a month". The EAI-produced spreadsheet models are shown in Figure 2.

The prompt for a month (Problem Statement 6) caused the GPT to generate one row for each day of the current month. It generated a series of daily prices for apple, with two decimal points (which is appropriate for a price), with values that are not unreasonable. It generated a series of daily numbers of apples bought, with integer values (which is appropriate), ranging from 1 to 13 (which is reasonable). There is much to be impressed by! However, our standards are high and they were not met.

Problem Statement 6 is an interesting failure. In a narrow sense it is accurate, in that column A: D contain correct values. However, column D contains numbers, not cell formulas which violates ERFR and means the spreadsheet model is not reusable. In addition, column E contains a garbage formula that calculates nothing of interest. Upon closer inspection, the formula in column E is analog of what the column D cell formula should be, but offset by one column.

In addition, we made additional EAI runs for Problem Statement 6. These runs returned different spreadsheets, indicating that EAI runs are not necessarily reproducible. We will see this again in section 4 and discuss it in section 6.

Figure 2: Spreadsheet Outputs for Problem Statements 4 (left) and 6 (center and right).

# 4. Experiments on a Larger Problem Statement (The Wall Task)

The Wall Task problem statement (in Appendix 4) appeared in Panko 1999 where it was used to investigate student performance on a modest spreadsheet modeling task. It has since been used by faculty at least a few universities as an exercise for students that connects them to the research literature. Because our students are familiar with it, it is a congenial candidate for the GPT.

We performed a handful of experiments, and found them to be informative, sometimes delightful but sometimes troubling. We used EAI on a prompt consisting of the Wall Problem Statement plus the three instructions ("Create an Excel model. Use cell formulas. Provide a downloadable Excel file."). The result was a spreadsheet with correct numbers, but no cell formulas (Figure 3). The GPT provided an option to "Make it a dynamic calculator with input fields", which we selected. This resulted (Figure 4) in a lovely model that satisfied ERFR with a bonus of distinct modules for inputs (first two columns) and outputs (last two columns).

| Material | Volume (cu ft) | Labor Cost ($) | Material Cost ($) | Total Cost ($) | Bid Price ($) |
|---|---|---|---|---|---|
| Lava Rock | 240 | 576 | 720 | 1296 | 1684.8 |
| Brick | 240 | 576 | 480 | 1056 | 1372.8 |

Figure 3:  Correct Output, No Cell Formulas

| INPUTS | | OUTPUTS | |
|---|---|---|---|
| Wall Length (ft) | 20 | Volume (cu ft) | 240 |
| Wall Height (ft) | 6 | Labor Cost ($) | 576 |
| Wall Thickness (ft) | 2 | Lava Rock Material Cost ($) | 720 |
| Crew Members | 2 | Brick Material Cost ($) | 480 |
| Hours per Day | 8 | Total Cost - Lava Rock ($) | 1296 |
| Work Days | 3 | Total Cost - Brick ($) | 1056 |
| Hourly Wage ($) | 10 | Bid Price - Lava Rock ($) | 1684.8 |
| Benefits Multiplier | 1.2 | Bid Price - Brick ($) | 1372.8 |
| Lava Rock Cost ($/cu ft) | 3 | | |
| Brick Cost ($/cu ft) | 2 | | |
| Profit Margin | 0.3 | | |

| | A | B | C | D |
|---|---|---|---|---|
| 1 | INPUTS | | OUTPUTS | |
| 2 | Wall Length (ft) | 20 | Volume (cu ft) | =B2*B3*B4 |
| 3 | Wall Height (ft) | 6 | Labor Cost ($) | =B5*B6*B7*B8*B9 |
| 4 | Wall Thickness (ft) | 2 | Lava Rock Material Cost ($) | =D2*B10 |
| 5 | Crew Members | 2 | Brick Material Cost ($) | =D2*B11 |
| 6 | Hours per Day | 8 | Total Cost - Lava Rock ($) | =D3+D4 |
| 7 | Work Days | 3 | Total Cost - Brick ($) | =D3+D5 |
| 8 | Hourly Wage ($) | 10 | Bid Price - Lava Rock ($) | =D6*(1+B12) |
| 9 | Benefits Multiplier | 1.2 | Bid Price - Brick ($) | =D7*(1+B12) |
| 10 | Lava Rock Cost ($/cu ft) | 3 | | |
| 11 | Brick Cost ($/cu ft) | 2 | | |
| 12 | Profit Margin | 0.3 | | |

Figure 4:  ERFR Spreadsheet Output, Providing Inputs and Outputs Modules. (Top shows values, bottom shows numbers.)

This is a curious thing; the prompt to use cell formulas yielded a spreadsheet without cell formulas, but then the follow-on prompt built a perfect spreadsheet model. We use The Wall Task as an in-class "personal benchmarking" exercise on the first day of our spreadsheet analytics course; had a student turned in this spreadsheet as their answer we would have congratulated them on their skill and near-perfect industrial-quality analytical spreadsheet model.

This motivated us to use EAI on a prompt of the Wall Task Problem Statement with no instructions. The resulting spreadsheet model (Figure 5) is functionally identical to Figure 4 (no cell formulas, identical numbers) but with different formatting and with the addition of a title.

**Wall Construction Bid Estimate**

| Wall Type | Volume (cu ft) | Labor Cost | Material Cost | Total Cost (No Profit) | Bid Price (with 30% profit) |
|---|---|---|---|---|---|
| Lava Rock | 240 | 576 | 720 | 1296 | 1684.8 |
| Brick | 240 | 576 | 480 | 1056 | 1372.8 |

Figure 5:  Correct Output, No Cell Formulas

Again, there was an option to provide a "dynamic" spreadsheet model. We selected this option, and EAI generated a spreadsheet file that generated an error message on opening requiring Excel to "recover" the contents. The resulting spreadsheet (Figure 6) has cell formulas some of which are accurate, some inaccurate, and some missing. It cannot be used and it is questionable whether it is useful as a "rough draft".

|   | A | B | C | D | E | F |
|---|---|---|---|---|---|---|
| 1 | INPUTS | | | | | |
| 2 | Wall Length (ft) | 20 | | | | |
| 3 | Wall Height (ft) | 6 | | | | |
| 4 | Wall Thickness (ft) | 2 | | | | |
| 5 | Hourly Wage ($) | 10 | | | | |
| 6 | Crew Size | 2 | | | | |
| 7 | Work Days | 3 | | | | |
| 8 | Hours per Day | 8 | | | | |
| 9 | Benefit Rate (%) | 20 | | | | |
| 10 | Lava Rock Cost ($/cu ft) | 3 | | | | |
| 11 | Brick Cost ($/cu ft) | 2 | | | | |
| 12 | Profit Margin (%) | 30 | | | | |
| 13 | | | | | | |
| 14 | | | | | | |
| 15 | Wall Type | Volume (cu ft) | Labor Cost | Material Cost | Total Cost (No Profit) | Bid Price (with Profit) |
| 16 | Lava Rock | =B2*B3*B4 | | =B12*B2*B3*B4 | =C16+D16 | =E16*(1+B13/100) |
| 17 | Brick | =B2*B3*B4 | | =B11*B2*B3*B4 | =C17+D17 | =E17*(1+B13/100) |

Figure 6: Spreadsheet Output Providing Inputs and (Unlabeled) Outputs Modules. Cell Formulas Are Correct, Incorrect, and Missing.

It appears that using the three instructions is helpful. But their performance is hard to understand. Further research is needed to try to better understand what is going on here.

## 5. WHAT MIGHT BE THE VALUE OF A GPT FOR BUILDING A SPREADSHEET MODEL?

Let's suppose that a GPT has provided a spreadsheet model. How can we think about its value? If our expectation is that the GPT has produced a satisfactory spreadsheet model that satisfies ERFR, how can we confirm this expectation? We call this the "problem of confidence".

If the spreadsheet turns out not to be accurate, or if our expectation is that the GPT has provided a useful "first draft" of a spreadsheet model (where "first draft" implies there are problems or at least imperfections), then how do we proceed to transform it into a model that satisfies ERFR? We call this the "problem of workflow".

### 5.1 The Problem of Confidence

How can we have confidence in a spreadsheet that is provided to us by an external agent? That agent might be a colleague, or a student, or nowadays a GPT. In particular, we want to be confident that the spreadsheet meets ERFR (each input in a cell; cell formulas; no hardwired numbers; labels; accurate). The first four can be trivially verified using View Formulas in Excel.

But the final requirement (accuracy) is harder to verify. How can we know whether the calculations are correct? This is a significant problem independent of the provenance of the spreadsheet. Panko 2006 and section 3.5 of Powell and Baker 2017 provide guidance on verifying accuracy. There are two approaches to confirming accuracy: testing and code examination.

In practice, analytical spreadsheet models are built to address a problem that has not previously been modeled. Consequently (section 7 of Grossman and Özlük 2004) the generation of test cases is a significant barrier, since they usually do not exist. Although trivial tests (such as setting all inputs to 0 or to 1) can identify certain errors, they cannot verify accuracy.

Code examination can be informal or formal, and Panko 1999 argues that multi-person code inspection has the highest error-detection rate. Spreadsheet auditing tools can automate the search for certain errors. I.e., verifying accuracy is difficult.

Fortunately, in this research we can quickly verify accuracy, because we possess accurate spreadsheets that allow instant creation of test cases. In addition, the models are small enough that single-person code examination allows evaluation of accuracy with a minute or two of effort.

## 5.2 The Problem of Workflow

In some fields of business, people argue that using a GPT to generate a "rough draft" that contains imperfections or even outright errors is a good idea. For example, writing (and also responding to) requests for proposals (RFP) is an area with much activity and several viable businesses (e.g., rohirrim.ai).

The argument boils down to workflow: starting with a GPT rough draft allows one to more quickly develop a high-quality final product, makes errors of omission less likely, and increases productivity. This is grounded in the logic of technology adoption: when a new technology arrives, it is wise to change one's business processes to leverage the technology's capabilities in order to achieve desired outcomes; old problems are reduced, new problems emerge, and a new workflow evolves. This obviously requires effort, learning, disruption, and a willingness to release past practices. And it doesn't always work.

It seems wise to explore this "rough draft thinking" to analytical spreadsheet models. Which raises the question, what is one to do with an imperfect spreadsheet model? One can study it for insight, then start over. Or one can build on and try to correct any imperfections or errors. This is not a trivial question and one that is not well understood.

The answer will undoubtedly differ by use case. Some use cases do not seem fertile ground at this time, for example high-stakes, large-scale financial planning models (surveyed in Grossman and Özlük 2010). But perhaps there are use cases where this does make sense, especially for less-sophisticated modelers.

We see two possible value propositions. First is total time. Second is the level of personnel skill. Total time is the time spent on prompting the GPT, plus the time spent on understanding and evaluating the resulting rough draft spreadsheet, plus the time spent on refactoring the rough draft spreadsheet into a final product (or starting over after being informed and inspired by the rough draft). If this total time is less than the time needed to code a spreadsheet from scratch, then the rough draft would add value.

The level of personnel skill is the combination of domain knowledge, spreadsheet knowledge, and (for organizations with spreadsheet standards) knowledge of the design standards and ability to quickly and reliably meet the standard. If it is possible for personnel with a lower level of skill (and hence pay) to develop a viable spreadsheet, then the rough draft would add value.

This seems an area of fruitful research. There is much activity in this space by many different players, and we anticipate that viable use cases will be emerging in the next few years.

# 6. CONCLUSIONS

This paper provides a summary of simple, entry-level experiments into using GPTs to build spreadsheet models. There is a large space of GPTs that claim an ability to build spreadsheet models. We found that the tools are temperamental and unreliable. However, they sometimes work very well, although the results are often not reproducible. Initial screening suggested that one tool (the ChatGPT custom GPT Excel AI by pulsrai.com) was a good choice.

We find it useful to think of analytic prompts as "problem statement" plus "instructions", where the problem statement can be given to any analyst or software system, and the instructions are specific to the situation. We explored different instruction sets for EAI, and landed on three simple sentences that were effective for our small problem sets.

We did experiments with 8 problem statements to examine three questions, and found that EAI is able to model parameterized problem statements, can model a month as 30 days, and can handle made-up nouns.

Because results are erratic, we find that the user must have sufficient skill to be able to evaluate the details of the spreadsheet and cell formulas, so the GPTs are not suitable for novice spreadsheet modelers.

**Experimentation is Difficult**

There are sometimes problems with reproducibility – the same prompt results in different spreadsheet models. Sometimes the differences are minor, sometimes they are problematic. Regardless, the inability to reproduce results makes systematic analysis more difficult.

In addition, the GPTs are a moving target. During the three-month course of this project, the performance and output of EAI was changing, presumably due to improvement efforts by the vendor. Comparisons need to be made at the same time to have hope of consistency across runs.

**Not Ready for Prime Time**

We anticipate that GPTs in this area will continue to grow and evolve. However, at this time, the immediate utility for GPTs to build spreadsheet models from problem statements of practical value is poor.

## 6.1 Future Research

There are many opportunities to extend this research.

The devisement of instructions is an open question. There is much wisdom on pre-prompting and prompt engineering for semantic and coding models that might prove relevant.

Most of our problem statements are simple calculations, at the level of "starter tasks" for beginner students of Excel. We briefly explored a single-paragraph problem statement (The Wall Task). A large corpus of modest business-related problem statements can be found as examples and homework problems in spreadsheet business analytics textbooks. Any fully satisfactory spreadsheet modeling GPT should be able to handle most or all of the problems in such a corpus.

The challenge of consistency and reproducibility needs rigorous examination, holding constant all possible sources of friction, including multiple tabs, time of day, location of computer, etc.

The primary challenge remains the ability to create spreadsheet models that satisfy ERFR with a high degree of reliability, to eliminate or at least reduce the amount of human evaluation and intervention required.

## 8. APPENDICES

### Appendix 1 - Summary of Evaluation of five GPTs

In Tables 1a and 1b we provide a summary of the basic facts and our evaluation of the 5 GPTs. (Note: these tables were generated by ChatGPT based on our research notes.)

|  | **GPT Excel** | **Doc Maker** | **Document Maker** |
|---|---|---|---|
| **Provider** | By NAIF J ALOTAIBI | By aidocmaker.com | By community builder |
| **User Ratings** | Ratings (1K+) - 3.8/5.0 | Ratings (1K+) - 3.7/5.0 | Ratings (1K+) - 3.6/5.0 |
| **Conversations** | 100K+ conversations | 200K+ conversations | 100K+ conversations |
| **Conversation Quality** | Shortest and least descriptive. GPT Excel does not deeply explain the generated content and process. | Shorter and less descriptive. Doc Maker provides input contents but does not deeply explain the generated content and process. | Detailed and informative. Document Maker provides formula calculation logic and input contents but does not deeply explain the generated process. |
| **User Interaction** | Less interactive. Generates models with some dialogue and requests clarification from the user. | Least interactive. Generates models with minimal dialogue and rarely requests clarification from the user. | More interactive and adaptive. Asks many clarifying questions. |
| **Error Handling** | Fewer errors, sometimes missing formulas and title. | Many errors, two model results become "0", one with formula, one without formula. | Fewest errors, but contains complex formulas |
| **File Reliability** | Reliable in some cases, Issues with formatting and missing formulas. | Unreliable in some cases, Issues with formatting, formulas and wrong results. | Highly reliable, Consistent structure and usable files with formulas. |
| **Visual Structure** | Moderate structure, decent formatting but lack of model title and | Poorly structure, lack of model title and divided input and output session. | Moderate structure, decent formatting but lack of |

|  | GPT Excel | Doc Maker | Document Maker |
|---|---|---|---|
|  | clear "output" text and session. |  | model title and clear "output" text and session. |

Table 1a:  Summary for three GPTs, GPT Excel, Doc Maker, Document Maker

**Comparison of Conversations and Ratings:  Excel AI vs. Spreadsheet Maker**

|  | **Excel AI** | **Spreadsheet Maker** |
|---|---|---|
| **Provider** | By pulsr.co.uk | By aidocmaker.com |
| **User Ratings** | 4.0 / 5.0 based on 50K+ ratings | 3.3 / 5.0 based on 1K+ ratings |
| **Conversations** | 2M+ conversations | 100K+ conversations |
| **Conversation Quality** | Detailed and informative. Excel AI provides step-by-step explanations of the model, including the logic behind formulas. | Conversations are typically shorter and less descriptive. Often, it lacks detailed explanations about the model or its construction process. |
| **User Interaction** | Interactive and adaptive. Asks clarifying questions, refines outputs through user input, and provides clear guidance throughout. | Less interactive. Generates models with minimal dialogue and rarely requests clarification from the user. |
| **Error Handling** | Fewer errors, usually limited to the input sections. Mistakes are easier to identify and fix due to clearer structure and labeling. | More frequent errors, including unfinished formulas, wrong cell references, and unsupported file formats may result in empty or non-functional spreadsheets. |
| **File Reliability** | Generally reliable outputs with consistent formatting. Generated files are functional, with errors being minor and non-disruptive. | Unreliable in some cases. Files may have formatting issues, incorrect formulas, or may download as empty/unsupported spreadsheets. |
| **Visual Structure** | Well-organized spreadsheets. Inputs, calculations, and outputs are clearly separated and labeled, making the model easy to follow. | Poorly defined visual structures. Despite separating inputs and outputs, the overall layout lacks clarity and logical flow, making it difficult for users to navigate the model. |

Table 1b:  Summary for two GPTs, Excel Ai and Spreadsheet Maker

# Appendix 2 - Information on EAI

Almost no information about EAI is provided on the pulsrai.com website. We interrogated EAI by asking "what is excel AI please tell me a little bit about yourself and what you can claim you can do", and it provided the following information:

> 💡 What Am I (Technically Speaking)?
>
> I'm an AI-powered virtual assistant, specifically a cloud-based conversational agent (or AI model) designed to help with Excel tasks. More formally:
>
> 👉 I'm a Generative AI built on large language models (LLMs), customized for Excel productivity and data operations.
>
> 👉 I work outside your local environment (no installations needed).
>
> 👉 I process inputs (data, requests, files) and return outputs (Excel files, reports, code, charts).
>
> You might call me:
>
> ✅ A Cloud-Based AI Service
>
> ✅ Conversational AI / Digital Assistant
>
> ✅ Generative AI for Excel
>
> ✅ AI-Powered Data Assistant

(Note: repeating this query yields a broadly similar, but different response. Reproducibility is not a feature of this product.)

## Appendix 3 - Problem Statements for EAI

We present eight problem statements below. The first two are a very simple situation to illustrate performance. Problem Statements 3-8 are slightly more complex. (As discussed in section 3.2, the prompt corresponding to each problem statement is created by appending to the problem statement these instructions: "Create an Excel model. Use cell formulas. Provide a downloadable Excel file.")

Area, using Data
> 1) Calculate the square footage of a rectangular room that is 10 feet long and 12 feet wide.

Area, using Parameters
> 2) Calculate the square footage of a rectangular room given its length and width.

Spending, using Data
> 3) A man buys two apples every day for 30 consecutive days, spending $2 per apple. What is total spending?

Spending, using Parameters

4) A man buys apples every day for 30 consecutive days. Given the price per apple and the number of apples he buys daily, what is total spending?

Spending, using Data, using Month

5) A man buys two apples every day for a month, spending $2 per apple. What is total spending?

Spending, using Parameters; using Month

6) A man buys apples every day for a month. Given the price per apple and the number of apples he buys daily, what is total spending?

Spending, using Data; using Made-Up Noun

7) A man buys two snapplees every day for 30 consecutive days, spending $2 per snapplee. What is total spending?

Spending, using Parameters, using Made-Up Noun

8) A man buys snapplees every day for 30 consecutive days. Given the price per snapplee and the number of snapplees he buys daily, what is total spending?

## Appendix 4 – Problem Statement for the Wall Task

Suppose that you are working for a general contractor ("the original consulting firm") who has asked you to build a spreadsheet model to help her to create a client bid to build a wall. According to your boss, you will start by offering two options—lava rock or brick. Both walls will be built by crews of two. Crews will work three, eight-hour days to build either type of wall. The wall will be 20 feet long, 6 feet tall and 2 feet thick. Wages will be $10 per hour per person. You will have to add 20% to wages to cover benefits. Lava Rock will cost $3 per cubic foot. Brick will cost $2 per cubic foot. Your bid must add a profit margin of 30% to your expected cost.